\documentclass[sigconf]{acmart}
\acmSubmissionID{607}

\usepackage{booktabs} % For formal tables

\usepackage[ruled]{algorithm2e} % For algorithms

\SetAlFnt{\small}
\SetAlCapFnt{\small}
\SetAlCapNameFnt{\small}
\SetAlCapHSkip{0pt}

\copyrightyear{2024}
\acmYear{2024}
\setcopyright{acmlicensed}\acmConference[SIGGRAPH Conference Papers '24]{Special Interest Group on Computer Graphics and Interactive Techniques Conference Conference Papers '24}{July 27-August 1, 2024}{Denver, CO, USA}
\acmBooktitle{Special Interest Group on Computer Graphics and Interactive Techniques Conference Conference Papers '24 (SIGGRAPH Conference Papers '24), July 27-August 1, 2024, Denver, CO, USA}
\acmDOI{10.1145/3641519.3657447}
\acmISBN{979-8-4007-0525-0/24/07}

\usepackage{graphicx}
\usepackage{amsmath}
\usepackage{booktabs}
\usepackage{appendix}
\usepackage{subcaption}
\usepackage{bm}

\usepackage{listings}
\usepackage{color}

\definecolor{dkgreen}{rgb}{0,0.6,0}
\definecolor{gray}{rgb}{0.5,0.5,0.5}
\definecolor{mauve}{rgb}{0.58,0,0.82}

\usepackage[scaled]{beramono} 
\usepackage[T1]{fontenc}

\usepackage{enumitem}
\usepackage{tabularx}
\usepackage[capitalize]{cleveref}
\crefname{section}{Sec.}{Secs.}
\Crefname{section}{Section}{Sections}
\Crefname{table}{Table}{Tables}
\crefname{table}{Tab.}{Tabs.}

\usepackage{ifthen}

\newboolean{showComment}
\setboolean{showComment}{true} % Set to 'true' or 'false'

\ifthenelse{\boolean{showComment}}{
    
\newcommand{\purvi}[1]{\textcolor{blue}{[purvi]: #1}}
\newcommand{\KCW}[1]{\textcolor{blue}{[Jackson: #1]}}
\newcommand{\needcite}[1]{\textcolor{red}{[CITE: #1}}
\newcommand{\karen}[1]{\textcolor{red}{[Karen: #1]}}
\newcommand{\purvinew}[1]{\textcolor{blue}{[New: #1]}}
\newcommand{\jacksonnew}{\textcolor{blue}{[J New:] }}
\newcommand{\vishnu}[1]{\textcolor{blue}{[Vishnu]: #1}}

\newcommand{\example}[1]{\textcolor{black}{#1}}
\newcommand{\sofia}[1]{\textcolor{blue}{[Sofia: #1]}}

}{

\newcommand{\purvi}[1]{}
\newcommand{\KCW}[1]{}
\newcommand{\KF}[1]{}
\newcommand{\needcite}[1]{}
\newcommand{\karen}[1]{}
\newcommand{\purvinew}[1]{}
\newcommand{\jacksonnew}[1]{}
\newcommand{\vishnu}[1]{}
\newcommand{\example}[1]{#1}
\newcommand{\sofia}[1]{}
}

\newcommand{\Xsrc}{$\mathbf{X}_\text{S}$\xspace}
\newcommand{\Xedit}{$\mathbf{X}_\text{E}$\xspace}
\newcommand{\Xsctx}{$\mathbf{X}^{\scriptscriptstyle \text{ctx}}_{\scriptscriptstyle \text{S}}$}
\newcommand{\xkeyS}{$\mathbf{x}^{\scriptscriptstyle \text{key}}_{\scriptscriptstyle \text{S}}$}
\newcommand{\xkeyE}{$\mathbf{x}^{\scriptscriptstyle \text{key}}_{\scriptscriptstyle \text{E}}$}

\begin{document}

\title{Iterative Motion Editing with Natural Language}
% \title{Character Motion Compiler: Iterative Motion Editing With Natural Language}
% \title{Character Motion Compiler: Iterative Motion Editing With Natural Language Using LLMs}

% \author{Purvi Goel$^1$, \ Kuan-Chieh Wang$^{2}$, \ C. Karen Liu$^1$, \ Kayvon Fatahalian$^1$  \\
% $^1$Stanford University,$^2$Snap Inc. \\
% }
\author{Purvi Goel}
\orcid{1234-5678-9012-3456}
\affiliation{%
 \institution{Stanford University}
 \country{USA}
}
\email{pgoel2@cs.stanford.edu}
\author{Kuan-Chieh Wang}
\affiliation{%
 \institution{Snap, Inc.}
 \country{USA}
}
\email{jwang23@snapchat.com}
\author{C. Karen Liu}
\affiliation{%
 \institution{Stanford University}
 \country{USA}
}
\email{karenliu@cs.stanford.edu}
\author{Kayvon Fatahalian}
\affiliation{%
 \institution{Stanford University}
 \country{USA}
}
\email{kayvonf@cs.stanford.edu}

\begin{abstract}

Text-to-motion diffusion models can generate realistic animations from text prompts, but do not support fine-grained motion editing controls.
In this paper, we present a method for using natural language to iteratively specify local edits to existing character animations, a task that is common in most computer animation workflows.  
Our key idea is to represent a space of motion edits using a set of kinematic motion editing operators (MEOs) whose effects on the source motion is well-aligned with user expectations. We provide an algorithm that leverages pre-existing language models to translate textual descriptions of motion edits into source code for programs that define and execute sequences of MEOs on a source animation. We execute MEOs by first translating them into keyframe constraints, and then use diffusion-based motion models to generate output motions that respect these constraints.
Through a user study and quantitative evaluation, we demonstrate that our system can perform motion edits that respect the animator's editing intent, remain faithful to the original animation (it edits the original animation, but does not dramatically change it), and yield realistic character animation results.

\end{abstract}

\begin{teaserfigure}
\centerline{
\setlength\tabcolsep{6pt}
\renewcommand{\arraystretch}{0.5}
\begin{tabular}{cc}
\includegraphics[width=0.49\linewidth]{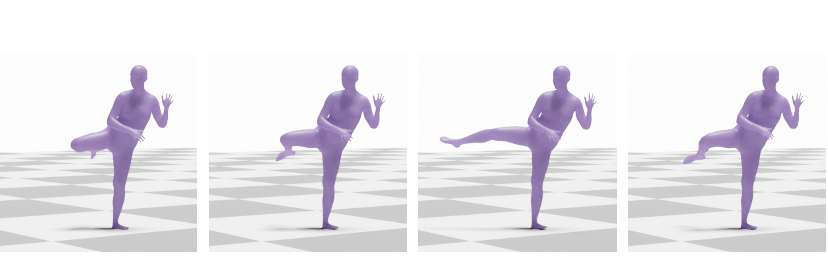} &
\includegraphics[width=0.49\linewidth]{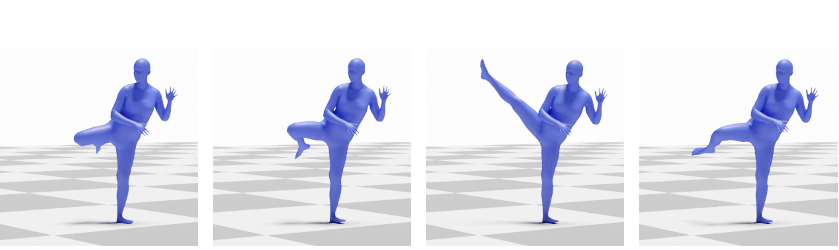} \\
Original Motion: \textit{A person is doing a side kick with the right leg} & Edit 1: \textit{Can you get that kick higher out?} \\
\includegraphics[width=0.49\linewidth]{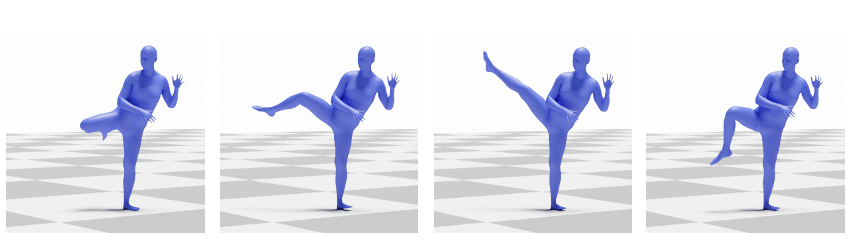} &
\includegraphics[width=0.49\linewidth]{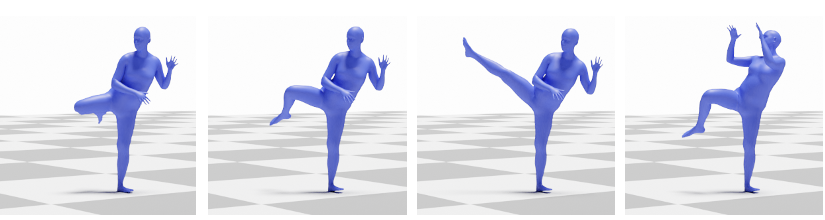} \\
Edit 2: \textit{Kick faster!} & Edit 3: \textit{After you kick, guard your face with your hands.}
\end{tabular}
}
\caption{ Our system supports iterative refinement of character motion using natural language. Here, the user has a vision for modifying an original kicking motion (top, left).   
Through a sequence of prompts, the user ``coaches'' the character to better match their artistic vision, progressively refining the motion by adjusting kinematic details.
First, the user requests the character to kick higher (edit 1), and then decides the kick should also be faster (edit 2). Finally, the user has the character raise its hands in  anticipation of a return attack (edit 3). 
Edited motions largely preserve the structure of the original motion while complying with the provided instructions. Retained conversation history allows the system to build upon previous edits.
}
\label{fig:banner}
\end{teaserfigure}

\begin{CCSXML}
<ccs2012>
<concept>
<concept_id>10010147.10010371.10010352</concept_id>
<concept_desc>Computing methodologies~Animation</concept_desc>
<concept_significance>500</concept_significance>
</concept>
</ccs2012>
\end{CCSXML}

\ccsdesc[500]{Computing methodologies~Animation}

\keywords{Character animation, motion editing, large language models, motion diffusion.}

\maketitle

%%%%%%%%% ABSTRACT
% \input{abstract}

%%%%%%%%% BODY TEXT

\label{sec:intro}
\section{Introduction}

A common task in character animation workflow is to edit existing animation sequences to match a particular creative vision.  For example, \example{an animator might start with a martial arts kicking sequence downloaded from a stock animation library (or generated by a modern generative model, or estimated from video using recent human motion reconstruction techniques), and wish to make the character kick higher or raise their arms to guard their face during the kick.}

In most traditional animation systems, performing these precise edits requires tedious keyframing of joints.
Conversely, emerging systems based on generative text-to-motion models require only modifying an input prompt, e.g., changing the prompt \example{``a side kick'' to ``a high side kick''}. However, it can be hard to predict how these systems will interpret a prompt, and animators have no guarantee the modified motion will retain any correspondence with the original. Some recent generative models offer limited editing control, but require additional multi-modal input, such as dense target joint trajectories or editing masks~\cite{karunratanakul2023dno, shafir2023human, guo2023momask}, rather than a simple text-based interface.

In this paper, we seek the best of both worlds: precise editing control, delivered via an accessible text-based interface. As shown in Figure~\ref{fig:banner}, given an existing character animation sequence, we hope to allow an animator to use natural language to \emph{engage in a conversation} with the animation system, specifying precise edits that control changes to the character's motion at specific points in the animation (\example{``after you kick, guard your face with your hands''}). By iteratively refining the motion over the course of the conversation, we aim to allow an animator to produce a modified character animation that matches their artistic vision.   

An important principle underlying the design of effective creative tools is predictability\,\cite{maneesh:blackboxes:2023}. An animator should be able to build a mental model of what a system will do in response to a control input. Inspired by this principle, our key idea is to constrain the space of animation edits to a small set of kinematic motion editing operators (MEOs) that are sufficiently simple that their effects on the source motion are well-aligned with user expectations.
For example, MEOs can express constraints on a pose at a particular time (``\example{wrist joint in front of head}'') or specify that a segment of motion should be slowed down or sped up.
As a result, MEOs can be robustly translated into low-level joint edits that yield realistic output. Further, the edited motion is likely to be consistent with user expectations. 
At the same time, MEOs raise the level of abstraction of motion editing commands from keyframes to calls to programmatic operations. This makes it feasible to leverage LLM-based program synthesis techniques to automatically translate a natural language prompt, which may contain imprecise or ambiguous motion editing descriptions, into an executable program that makes API calls to create valid sequences of MEOs. Specifically we make the following contributions:

\begin{itemize}
    \item We propose a set of \textbf{kinematic motion editing operators} (MEOs) that express fine-grained control similar to keyframes, but present a higher level of editing abstraction by modeling edits as spatial and temporal constraints expressed relative to poses (or events) in the source motion (e.g., a hand \emph{above} a head, or a pose-change \emph{after} a foot contact). MEOs serve as a useful intermediate representation for bridging high-level motion editing intent and low-level motion editing operations in an iterative editing context. 
    \item We provide an algorithm, based on using LLMs for program synthesis, that translates natural language motion editing directions into \emph{Python programs} that consist of MEOs.
    \item We provide an algorithm for applying motion edits described by MEOs to a source character animation sequence. Our approach translates MEOs into keyframes that constrain the output motion, and leverages diffusion-based generative motion models to modify the source motion to adhere to these constraints while maintaining realistic human motion.  
\end{itemize}

Through qualitative and quantitative evaluation and a user study we demonstrate that our system provides an intuitive natural language interface for iterative character motion editing. Our system supports a  range of motion edits, and produces motions that are visually realistic, respect the intent of the user, and preserve the fundamental structure of the original motion. Code and data for this paper can be found on our project webpage.

\section{Overview and Design Goals}
\label{sec:goals}

Our goal is to support iterative editing of an existing character animation sequence.  Our system takes as input a starting motion \Xsrc (the character's root position and joint angles for each frame in the sequence), a plain-text description of that motion ($E_{\text{ctx}}$), and a plain-text editing instruction $E$. It generates an edited motion \Xedit adhering to the following desiderata:

\begin{enumerate}
\item \textit{High edit fidelity}. \Xedit should reflect the intended edit $E$. For example, if the edit is \example{``after you kick, guard your face with your hands''} (Figure~\ref{fig:banner}), the character's hands should be up after the kick in the animation \Xedit (but not at the beginning).

\item \textit{Non destructive}. \Xedit should minimally change aspects of the motion that should not be impacted by $E$. For example, \example{in the above example of a kick}, ideally the character's kick would be minimally impacted by raising the hands. 

\item \textit{Realistic}. \Xedit should be globally harmonized, meaning that the result should be a plausible character motion.

\end{enumerate}

Our goals can conflict in complex ways.  For example, to preserve realism when adding a \example{higher kick, it might be necessary to add additional transitional movement}, running against the goal of being as non-destructive to the original sequence as possible. 

After producing \Xedit, if the user has not achieved the desired motion, they may continue to iterate, repeating the process using a prior \Xedit as \Xsrc, and providing a new $E$. \example{For the example in Figure~\ref{fig:banner}, the iterative process involves editing the last generated \Xedit; for other editing scenarios, it may also involve the user backing up to a prior step in the session and modifying that motion instead.}

\section{Related Work}
\label{sec:related}

\textbf{Human motion editing} is a well-studied and challenging problem. Early work explored modifying motions with spacetime constraints~\cite{gleicher97, mopathedit, lee_edit}, but producing realistic, coordinated motion edits typically requires the user to manually provide dense constraints. %We adopt the style of specifying edits with spacetime constraints (defined using MEOs), but provide system that provides a natural language interface for specifying these constraints.
In recent years, deep learning (DL) has explored automating the process of motion editing with sparser user input. Key approaches include motion stylization~\cite{aberman2020unpaired, 10030800}, pose editing~\cite{oreshkin2022protores}, and in-betweening~\cite{tseng2022edge, tevet2023human, shafir2023human, twostagetransformer}. However, the above approaches do not use text-based control. Works like~\cite{Delmas_2023_ICCV} keep the natural-language interface we desire, but do not extend to motion.

%has been effective across several character animation domains: joystick control~\cite{motionvae}, action class~\cite{a2m, petrovich21actor, cervantes2022implicit}, learned skills~\cite{tessler2023calm, 2022-TOG-ASE,2022-SA-PADL}, and text~\cite{zhang2022motiondiffuse, zhang2023generating, petrovich22temos, Ahuja2019Language2PoseNL, SINC:ICCV:2022}. Key approaches to DL-based motion editing include motion stylization~\cite{aberman2020unpaired, 10030800} and pose editing/correction~\cite{Delmas_2023_ICCV, fieraru, oreshkin2022protores:}. While motion stylization techniques can perform global edits on motions, they do not enable precise control. Works like~\cite{Delmas_2023_ICCV} keep the natural-language interface and precise control we desire, but do not extend to motion.

\textbf{Programmatic representations of human motion} are a long-standing way to summarize motion sequences, such as with smaller clips or motifs~\cite{kovar,deepmotif}, learned concepts~\cite{endo2023humanmotionqa}, or combinations of primitives~\cite{motion2prog2021}. Unlike these works, we propose an intermediate representation (IR) that is specific to motion editing. In that vein, our representation is similar to that of~\cite{coaching}, which introduces an IR to edit dynamic controllers in a physics-based setting, but we focus our IR on text-driven~\textit{kinematic} motion edits.

\textbf{Plans as programs} is a strategy that represents problem solving plans as code. Our design of MEOs is inspired by recent systems that perform advanced reasoning about visual environments by using LLMs to translate high-level, plain-text problem descriptions into \emph{executable programs} that make calls to a pre-defined, domain-specific API to carry out a precise reasoning strategy. This strategy has been used for task planning in virtual environments~\cite{huang:2022:zeroshotplan,singh:2023:progprompt,liang:2023:codeaspolicies,wang:2023:voyager} and for visual question answering~\cite{surismenon:2023:vipergpt}.
By representing plans as executable code, these approaches simultaneously leverage the common sense reasoning and program synthesis capabilities of LLMs, and ground plans in the environment by providing APIs for the resulting programs to query for environment-specific information (e.g., nearby objects). We follow a similar design for expressing high-level motion editing intent and grounding edits in a target animation sequence.

% ChatGPT-4~\cite{brown2020language}, having remarkable abilities for program synthesis and task planning~\cite{wang2023voyager,huang2022language,ahn2022i,liu2023llmp}, are a well-suited tool. 

\textbf{Motion diffusion models} can generate high-quality 3D motion from text~\cite{tevet2023human, zhang2023finemogen, zhang2022motiondiffuse, ren2023insactor}.  
In addition to generation, motion diffusion models can learn a strong prior for tasks like constrained trajectory infilling~\cite{li2023object, rempeluo2023tracepace}, motion reconstruction from video~\cite{jiang2024back}, and multi-person reconstruction~\cite{muller2023generative}.
The prior makes the produced motions more realistic and plausible.  
Our work incorporates this strong motion prior to ensure the realism of the edited motion. 
While there is a plethora of diffusion-based editing methods in the image domain~\cite{brooks2022instructpix2pix,hertz2022prompt,sarukkai2023collage, meng2022sdedit}, they are not applicable as they depend on the architecture of the diffusion model.  
Specifically, they rely on manipulation of the cross attention layers which interface the input text encoder. 
The widely used text-conditioned MDM~\citep{tevet2023human} does not have this architecture. 
For the task of motion editing, work like~\cite{shafir2023human} modifies MDM's motion inpainting process to control and edit end-effector trajectories, but requires dense multi-modal input (e.g, the entire joint trajectory).~\cite{karunratanakul2023dno} similarly supports edits to motion trajectories, but does not support fine-grained edits nor a text-based interface for specifying corrections.

% \textbf{Diffusion Models.}~\cite{dickstein, ho2020denoising} have shown promise for generative modeling of images~\cite{Saharia2022PhotorealisticTD, rombach2021highresolution, radm} and video~\cite{blattmann2023videoldm, ho2022imagen}. \kf{recommend dropping prior sentence, it is not relevant to paper. Directly start with diffusion for motion, and maybe call paragraph "Diffusion-Based Motion Generation" or something like that.} They have also been applied to human motion generation using various forms of conditionings~\cite{tevet2023human, rempeluo2023tracepace, tseng2022edge}. Like Wei et al.~\cite{Wei2023UnderstandingTM}, we use keyframe conditioning, but introduce a different loss formulation and design choices more suited to our goals. Approaches for editing images using diffusion models~\cite{brooks2022instructpix2pix,hertz2022prompt,sarukkai2023collage, meng2022sdedit} is difficult to directly adapt to the motion domain; both ~\cite{brooks2022instructpix2pix,hertz2022prompt}, for example, rely on an image foundation model's per-word attention masks; SOTA motion diffusion models collapse text input into a single token. \cite{shafir2023human} modifies MDM's~\cite{tevet2023human} motion inpainting process to control end-effector trajectories, but require dense multi-modal input (e.g, the entire joint trajectory) as opposed to our text instructions.

\section{Method}
\label{sec:method}

\begin{figure}
    \centering
    \includegraphics[page=2,trim={80 600 80 50},clip,width=0.95\linewidth]{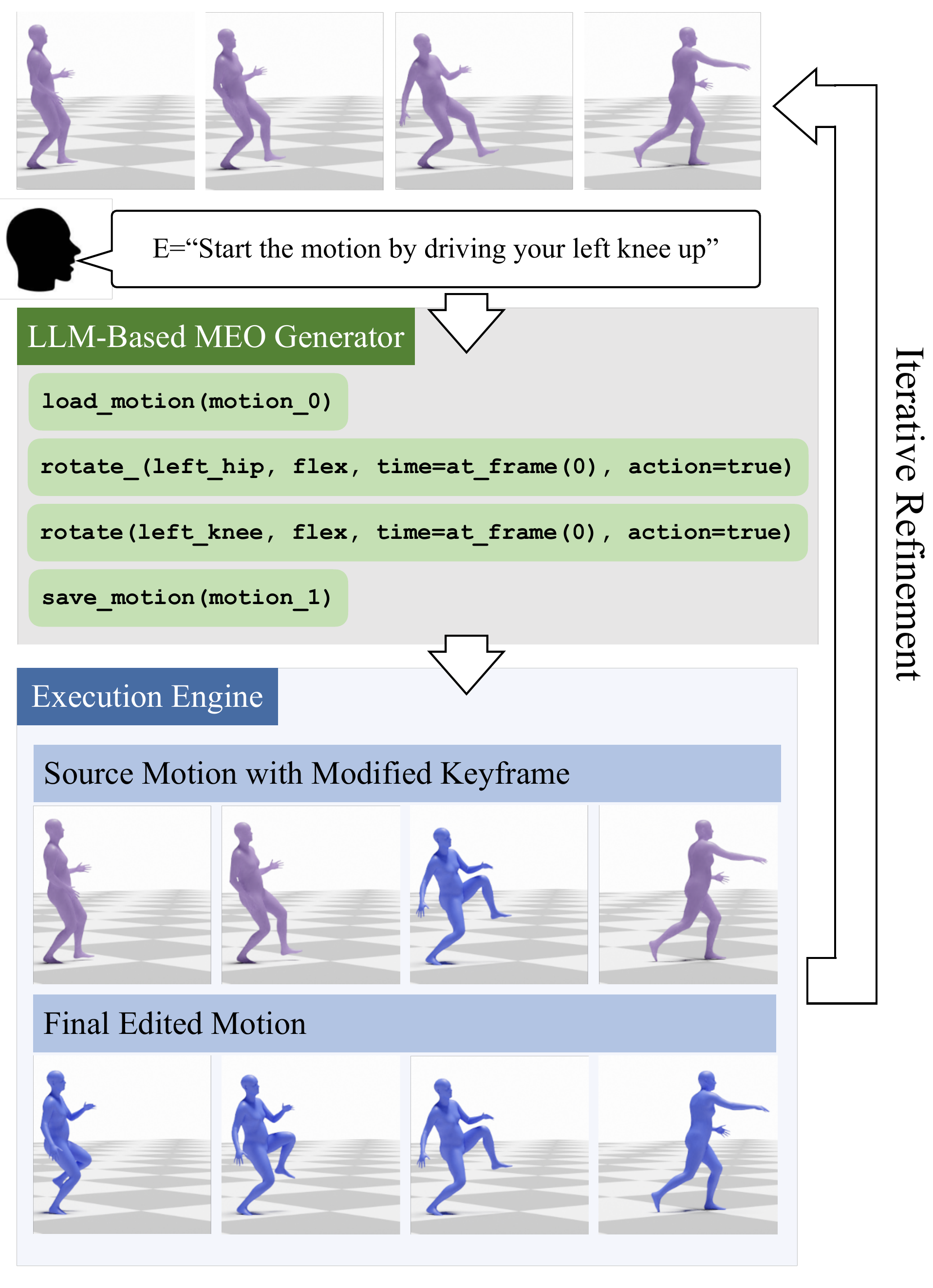}
    \caption{\textbf{System overview}: Our system uses a LLM to translate a natural language editing instruction ($E$) into source code for a Python program that executes motion editing operations (MEOs).
    Our MEO execution engine applies MEOs to the source motion by first generating motion constraints (e.g., keyframes, retiming constraints). In the case shown above, \textit{E} describes a sub-movement that should \textit{start} at the beginning of the motion and lead to a pose in the future; the engine determines the explicit frame requiring editing. A diffusion-based motion infilling step then produces output motions that embody the desired edit, preserve the original motion when possible, and look realistic. Our system can be used in an iterative fashion.} 
    \label{system}
\end{figure}

 % Our key idea, inspired by recent work utilizing LLMs in robot planning, is to cast motion editing as a two-step process: converting natural language editing instructions into Python programs that describe fine-grained editing operations, then executing resulting operations using a keyframe generation and diffusion-based motion infilling process.

The key idea of our approach, illustrated in Figure~\ref{system}, is to cast motion editing as a two-step process: first, converting natural language editing instructions into a sequence of discrete motion editing operations (MEOs), then executing resulting operations using a keyframe generation and diffusion-based motion infilling process.
% represent a high-level motion editing instruction ($E$) as sequences of motion editing operations (MEOs) that define kinematic edits.
We first describe the MEOs supported by the system.
(Section~\ref{sec:mep}). Then we describe how we use an LLM, prompted using in-context learning, to translate a plain-text motion editing instruction into an \emph{executable Python program} comprised of MEOs (Section \ref{sec:llm}). Finally we describe how we implement the motion edits described by MEOs to produce new motions (Section~\ref{sec:execution}).  

% During a motion editing session, the editing process is expected to be invoked repeatedly as a user iteratively refines an animation to meet their creative intent.

\subsection{Motion Editing Operators (MEOs)}
\label{sec:mep}

A common representation for kinematic motion specification is a keyframe, which defines the location or orientation of character's joints at a frame. 
Our system is inspired by traditional workflows for keyframe editing, but raises the level of abstraction with MEOs. 

Like keyframes, the majority of MEOs supported by our system define a joint to modify, a spatial constraint (rotation/translation) on that joint, and a time interval during which the constraint applies. 
Target joints can in principle be any of the joints in the SMPL model~\cite{Bogo:ECCV:2016}; we focus on the end effectors, knees, elbows, head, shoulders, hips, chest, and waist. 

\begin{figure}
    \centering
    \includegraphics[page=3,trim={0 550 0 50},clip,width=0.9\linewidth]{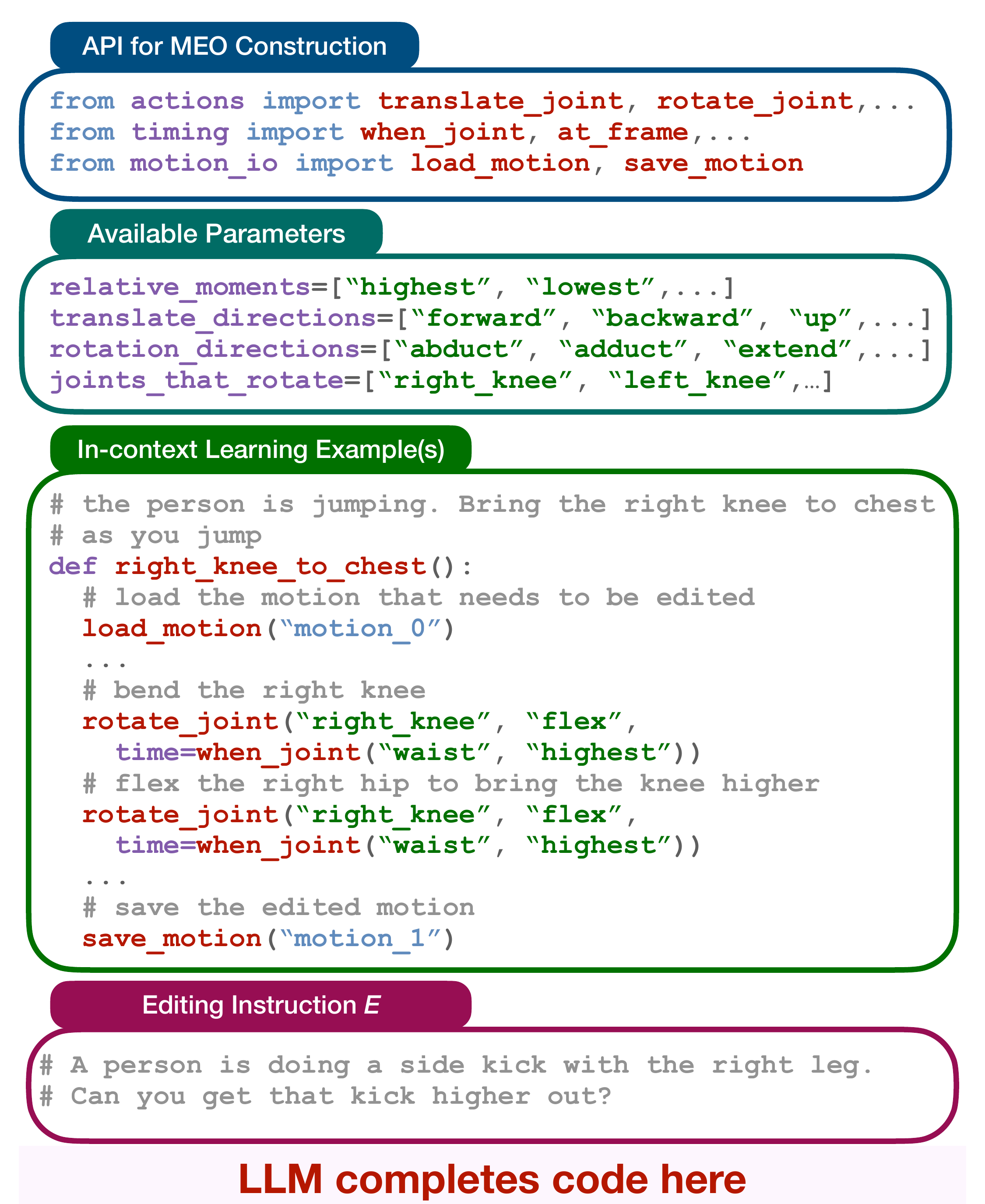}
    \caption{\textbf{LLM Prompt Specification}. An abridged LLM prompt that contains MEO API information, an editing prompt $E$: ``Can you get that kick higher out?'' (with context $E_{ctx}$ ``A person is doing a side kick with the right leg"), and an example MEO program for the task: ``lift the right knee to the chest during a jump.'', which serves to teach the LLM agent how to use the API. In practice, we provide several examples.  The example program here makes API calls to create a plan for completing the editing task, by using MEO construction methods from our API and lists of joints/directions.  We ask the LLM agent to write a program that performs the motion edit by combining $E$ and $E_{ctx}$ into a function header comment. The LLM completes the code by writing an MEO program under the header comment.}
    \label{fig:llm_code}
\end{figure}

Unlike keyframes, the spatial and temporal constraints of MEOs can be expressed \emph{relative} to properties of the source motion they are being applied to. Rotation/translation constraints may be defined relative to the joint's current configuration (e.g., moving the right hand \emph{higher}, or \emph{abducting} the left hip) or relative to another joint (e.g., moving the right hand \emph{above} the right shoulder). Spatial constraints are applied over a time range that is specified by temporal MEO parameters. Temporal MEO parameters can be explicit references to frame indices, e.g., \texttt{at\_frame}. We also include implicit references, e.g., \texttt{when\_joint}, to ground the MEO program in \Xsrc. For example, an MEO might reposition the right knee when the waist is highest. See Figs.~\ref{system} and~\ref{fig:llm_code} for examples. Additionally, MEOs designate if the edit describes a change in pose that should occur at the specified time, or describes a sub-movement that should begin at the stated time and lead to an edited pose in the future. 

To simplify the space of edits that must be accurately generated from natural language instructions, we limit translation and rotation spatial constraints to a small set of discrete directions (e.g, higher/lower, above/below, abduct/adduct), rather than specific numerical values or vectors (e.g, 10.2~cm). It is common for humans to use these coarse descriptions when talking about motion. Our system also supports non-keyframe-based MEOs, such as operations that speed up or slow down a segment of motion. 

We implement MEO abstractions as a Python API, containing methods for constructing MEOs and querying the source motion for specific times of extrema events. We provide full details of this API in the supplemental. An example of usage is given in Figure~\ref{fig:llm_code}.

\subsection{Generating MEOs from Natural Language}
\label{sec:llm}

Given a plain-text motion editing instruction~$E$, and a description of the source motion to modify, we prompt an LLM to generate Python using the MEO API to perform the editing task specified by $E$.  

\subsubsection{Context strings for grounding.} Interpreting a motion editing instruction 
requires understanding the context of \Xsrc.  For example, without knowledge of the contents of \Xsrc, it is unclear to the LLM which leg the instruction ``kick higher'' intends to modify. While LLMs are capable of handling some multi-modal tasks~\cite{feng2023posegpt, yan2021videogpt, gong2023multimodalgpt}, these models cannot yet interpret or produce 3D motion.
To address this grounding problem, in addition to the corrective motion editing instruction $E$ (e.g., \example{``kick higher''}) our system requires a motion context string $E_{\text{ctx}}$: a short description of the current motion (e.g. \example{``a person is doing a side kick with the right leg.''}). The latter can be provided by the user, automatic captioning~\cite{jiang2023motiongpt}, or, if the original motion was generated by a text-conditioned model, the original prompt.

\subsubsection{Prompt structure.} Like~\cite{singh:2023:progprompt}, we inform the LLM agent about the MEOs and time query functions available in our API via \texttt{import} statements, and provide the set of valid MEO parameters as a list of strings at the top of the file (e.g., valid joints, relative translation/rotation options). This has been shown to encourage the LLM to use only the methods and parameters it has available. In addition to these inputs we follow standard in-context learning practice and include a small collection of examples of valid MEO programs and their corresponding motion editing prompts~\cite{wei2023chainofthought}. These programs demonstrate how to use the MEO API functions and, in the case of iterative editing sessions (discussed below), how to access the correct motions in the motion undo stack and summarize a new $E_{\text{ctx}}$ from the session's history. 

At inference time, $E$ and $E_{\text{ctx}}$ are provided as code comments and the LLM agent is prompted to ``complete the code'' to satisfy $E$. When doing so, we ask the LLM to generate code comments that justify its choices of MEOs and MEO parameters. This form of self-reflection is known to improve the quality of LLM output~\cite{yao2023react, shinn2023reflexion}. See Fig~\ref{fig:llm_code} for an abridged example of a prompt containing one in-context example.
 If the generated program is invalid (e.g., the LLM uses a invalid function or parameter) the system reports the error message to the LLM agent, which tries generation again. 

% We observe that when describing motion edits in natural language, two forms of edits are common. Edits that are indented to statically re-pose character at a specified time (``Move the left hand higher at the start of the motion.'') and edits that are intended to describe a desired movement ("Kick your left leg up at the start of the motion.").  In both cases, 

% Every program begins with a function header comment, which is just instruction \textit{E}. Then, the motion to be edited is loaded from an in-memory cache by name, e.g., \texttt{load\_motion(``motion\_0")}. The function invokes a sequence of MEO operators as function calls and MEO parameters as arguments. For example, \example{``start by raising your arms out to the side"} includes the function call \texttt{do\_rotate(``abduct", ``right\_shoulder",time=at\_frame(0))}. Finally, the edited motion is saved to the in-memory cache, e.g, \texttt{save\_motion(``motion\_1")}.

% \subsection{Iterative Editing Support} 

\subsubsection{Iterative Editing Support.}
Editing motion is an iterative process, and a key goal in our system is to support iterative editing via an extended conversation between the user and the system. Iterative edits can be necessary to clarify or disambiguate the goal movement, or to break larger editing intents into sub-goals. For example, a human might correct, \example{``Kick higher'', ``Higher'', ``Now finish in a squat.''} Alone, edits like ``higher'' are ambiguous, but gain context from knowledge of previous instructions. So, we provide previous editing instructions and MEO program outputs from the entire session as part of the input prompt in each step. Thus the LLM agent can reference earlier conversation points without their explicit mention in a new $E$, can correct programs in the case of human-computer miscommunication, and build upon previous edits.

\subsubsection{Undo Stack.} During iterative editing sessions, the system must determine which motion an editing instruction refers to. For example, the instruction \example{``Keep your hand in front of your waist''}, followed by \example{``Now add a kick at the start''}, implies the second edit should modify the output of the first. Conversely, if the second instruction is \example{``No, your other hand''}, the edit should be applied to the original motion. Our runtime maintains a cache of motions produced during a session, and allows access to each using \texttt{load\_motion(``motion\_N'')} and \texttt{save\_motion(``motion\_N'')} API methods. Given a prompt \textit{E}, the LLM agent must choose which prior motion to load (what the parameter to \texttt{load\_motion()} should be); the next result is always saved as motion N + 1.

% single goal
\subsection{Execution Engine}
\label{sec:execution}
To execute the MEO progam, we need identify the specific frames \xkeyS to operate on. Then, we mechanically edit \xkeyS to produce edited frames \xkeyE. Finally, we integrate \xkeyE back into \Xsrc while retaining plausibility by leveraging the powerful generative prior of a diffusion model. Our motion notation is illustrated in Fig~\ref{notation}.

\subsubsection{Frame Identification}
We first identify the frame indices specified by each MEO. Explicit references require no processing of \Xsrc, but for implicit references, we analyze joint trajectories of \Xsrc and compute explicit frame indices using heuristics. For example, if the operation should occur \textit{when} a joint reaches an extremum, we identify the frame index containing the extremum.

\subsubsection{Spatial Constraints} Spatial edits are directly executed on \xkeyS to produce \xkeyE. If an MEO specifies a rotational or translational edit, we directly apply it with forward/inverse kinematics, respectively. Edit magnitudes are procedural, grounded in the current articulation of \xkeyS; for example, rotation edits take joints a fraction of the way from the current angle to the joint limit. We use spline-based time warping to change speed~\cite{Witkin1995MotionW}; the start/end of the identified time range act as time-warp constraints, manipulated programmatically based on the desired change. See our Supplemental for implementation details. 

\subsubsection{Generative Interpolation}
\begin{figure}
    \begin{center}
    \includegraphics[page=2,trim={150 450 600 190},clip,width=\linewidth]{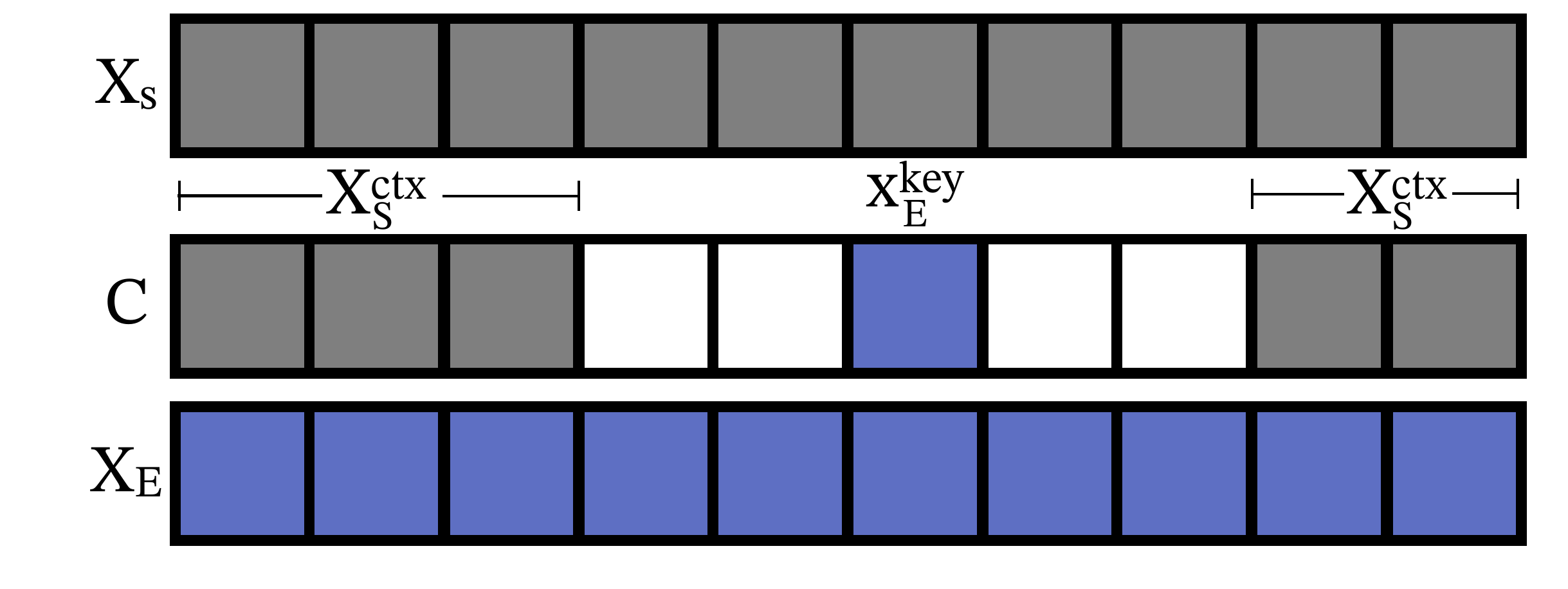}
    \end{center}
    \caption{\textbf{Motion notation}. \Xsrc is the source motion; condition \textbf{C} comprises \Xsctx (context from \Xsrc) and edited keyframe(s) \xkeyE. Our diffusion-based execution engine outputs \Xedit. Gray squares represent components of \Xsrc; blue squares represent components or desired components of \Xedit. }
    \label{notation}
\end{figure}

When new keyframes \xkeyE are added or edited, we need to modify neighboring frames such that transitions appear plausible. While traditional techniques like spline interpolation may to ensure motion smoothness, interpolated results may not appear natural. We cast the problem of integrating \xkeyE into \Xsrc as a motion-infilling problem, and leverage a motion diffusion model to solve for the transition.

Given the context frames \Xsctx and edited keyframe \xkeyE, our model needs to generate a completion of the motion, \Xedit. We extend diffusion models~\cite{ho2020denoising} thanks to their recent success.

The core component in diffusion models is a denoising network, $G$, trained to reverse the Markov noising process below:
\begin{equation}
q(\mathbf{X}_t | \mathbf{X}) = \mathcal{N}(\sqrt{\alpha_t}\mathbf{X}, (1-\alpha_t)I),
\end{equation}
where $\alpha_t \in(0, 1) $ decrease monotonically. We use a variant of the diffusion model that outputs the denoised motion at each step (the `simple' loss in~\cite{ho2020denoising}); denoiser $G$ takes as input the noised motion $\mathbf{X}_t$, the condition $\mathbf{C}$, and the current diffusion step $t$, and learns to output the denoised motion with objective:
\begin{equation}
\mathcal{L} = \mathbf{E}_{\mathbf{X}, t} [ \| \mathbf{X} - G(\mathbf{X}_{t}, \mathbf{C}, t) \|^2_2 ].
\end{equation}

The representation of $\mathbf{C}$ is a critical design choice. Since our goal is to teach the model to infill transitional motion around \xkeyE (which will be provided by MEOs at inference time) for a transition window of length \textit{W} and motion length $F$, we represent $\mathbf{C}$ as a motion sequence composed of context frames from the source \Xsctx: $\mathbf{x}^{\scriptscriptstyle 0:\text{key}-1-W}$ and $\mathbf{x}^{\scriptscriptstyle\text{key}+W:F-1}$, and with \xkeyE. Neighboring frames around \xkeyE, $\mathbf{x}^{\scriptscriptstyle\text{key}-W:\text{key}}$ and $\mathbf{x}^{\scriptscriptstyle\text{key}+1:\text{key}+W}$ are masked to zero (see Fig.~\ref{notation}).

During training, we randomly sample a keyframe index between $0$ and $F-1$ and zero out $W$ frames before/after the keyframe(s). The window is clamped at the start and end of the sequence. 

\subsubsection{A Diffusion-Based Architecture for Infilling Conditioning.} 
\begin{figure}
    \centering
    \includegraphics[page=4,trim={0 1000 0 200},clip,width=0.9\linewidth]{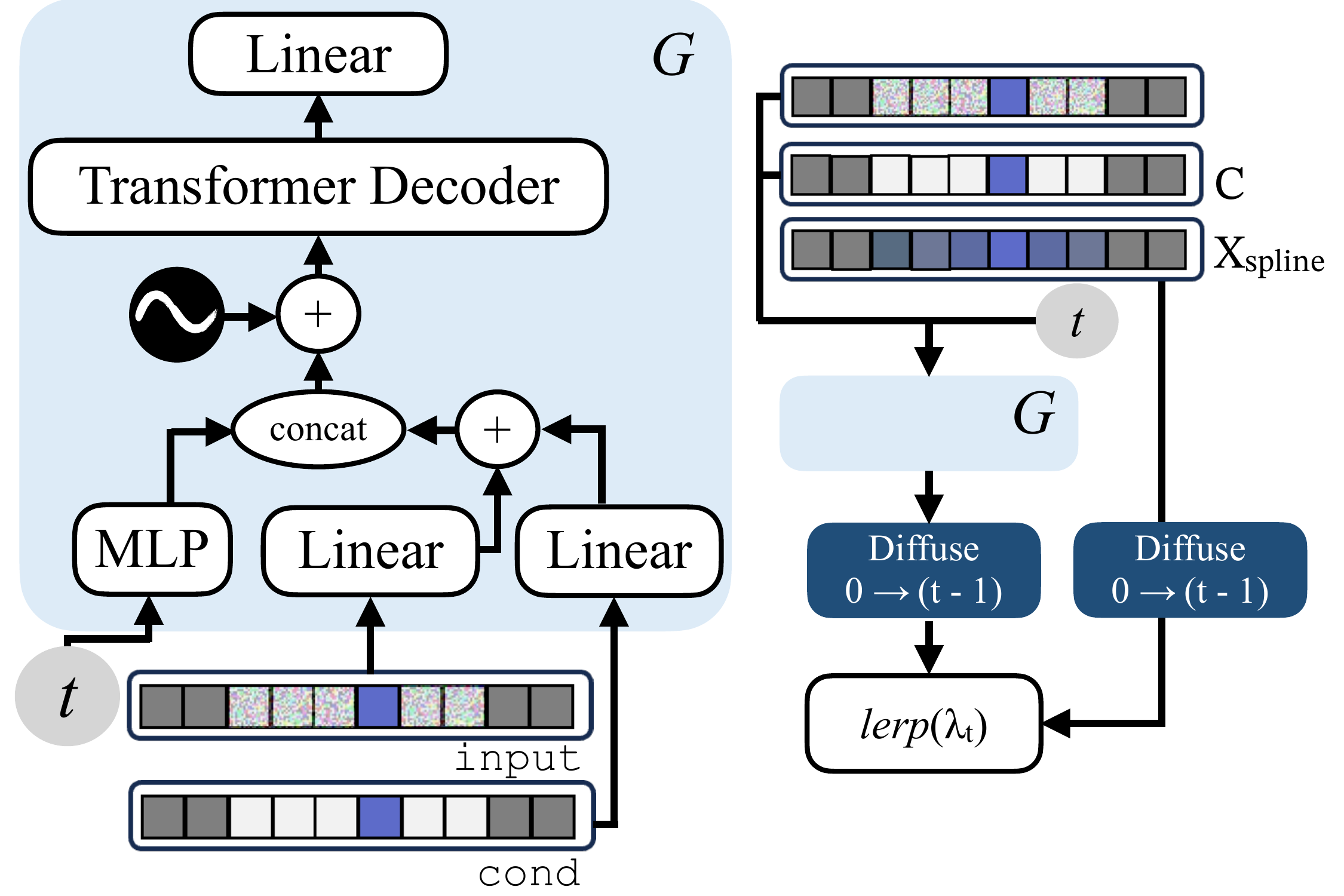}
    \caption{\textbf{Infilling Diffusion Model}. In training, our model \textit{(left)} learns to infill motions. \textit{G} takes \texttt{input}, a noisy sequence imputed with $\mathbf{C}$, and \texttt{cond}, a masked verion of $\mathbf{C}$. At inference \textit{(right)}, we optionally integrate $\mathbf{X}_{\text{spline}}$ to guide inference. For each $t$ we spatially \textit{lerp} the infilled frames of $\mathbf{X}_{\text{spline}}$ with those progressively generated by \textit{G} with interpolant $\lambda(t)$, which decreases monotonically as a function of $t$.}
    \label{fig:arch}
\end{figure}

We use an architecture based on the transformer-decoder, augmented with a conditioning branch to account for $\mathbf{C}$. To distinguish the original input branch from the conditioning branch, we use the following labels, `\texttt{input}' vs `\texttt{cond}' (Figure~\ref{fig:arch}).
In practice, since $\mathbf{C}$ can be thought of as a masked version of a full motion sequence $\mathbf{X}$, we introduce a binary mask $\mathbf{M}$ that zeros out attributes that need to be infilled. Like~\cite{Wei2023UnderstandingTM}, we feed context information in the original input branch rather than pure noise:
\begin{align}
G(\texttt{input=}&\mathbf{M} \odot \mathbf{X} + (1-\mathbf{M})\odot q(\mathbf{X}_t|\mathbf{X}),\nonumber \\ 
    \texttt{cond=}&\mathbf{M} \odot \mathbf{X}, t 
    ). 
\end{align}

\subsubsection{Inference.} 
At inference, \textbf{C} comprises \Xsctx and \xkeyE. $\mathbf{M}$ zeros out degrees of freedom that need to be infilled around \xkeyE. Similar to the image blending process proposed by \cite{Avrahami_2022_CVPR}, we observe more consistent generations for non-root edits when we seed the denoising process with an initial guess for the masked area; we use $\mathbf{X}_{\text{spline}}$, a naive solution for \Xedit which integrates \xkeyE into \Xsrc using spline interpolation. At each diffusion step \textit{t}, we spatially blend \textit(lerp) the infilled regions of an appropriately noised $\mathbf{X}_{\text{spline}}$ with $\mathbf{X}_t$ using monotonically decreasing interpolant $\lambda_{t}$. Our insight is that $\mathbf{X}_{\text{spline}}$ guides the start of the denoising process, while later diffusion steps add detail.

\subsubsection{Implementation.} 
In practice, we found that edits to the root joint, e.g., to make a character jump or crouch, were better handled if $\mathbf{C}$ included the root trajectory to help break the problem down. So, we trained two models in the manner described above: a regression model to infill the root trajectory, and \textit{G} to infill the rest of the body. At inference, the first model generates the root trajectory, which the second model includes in $\mathbf{C}$ to infill the other degrees of freedom. We train both models on the AMASS dataset \cite{AMASS:ICCV:2019}. 

Masking frames with $\mathbf{M}$ can destroy important structural information from \Xsrc. So, at inference, we automatically detect important frames in \Xsrc that should be preserved and include these in $\mathbf{C}$. Frames are important if they either contain significant extrema, or were edited in a previous iteration. \Xedit can optionally be postprocessed with, e.g., smoothing and foot-skate clean-up.

\section{Evaluation}
\label{sec:eval}

\subsection{Implementation Details}

We use OpenAI's ChatGPT-4 as our LLM agent. We train diffusion model \textit{G} and the trajectory infilling model on the AMASS dataset~\cite{AMASS:ICCV:2019} using an NVIDIA Tesla V100 GPU. All motions are represented as 60-frame clips (2.5 seconds). Hyperparameters are included in the Supplemental.

\begin{figure*}

\begin{subfigure}[t]{0.49\linewidth}
\includegraphics[width=\linewidth]{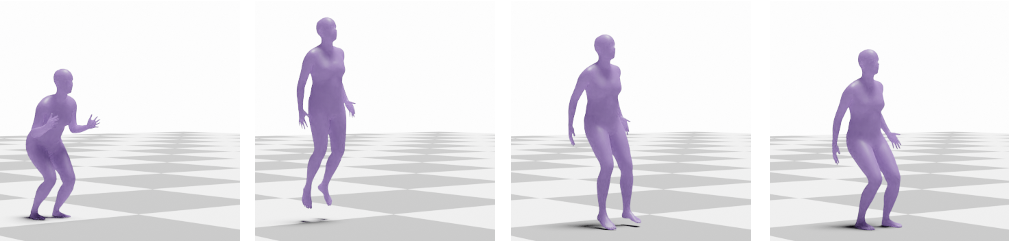}
\caption{Original Motion: \textit{A person is jumping.}}

\end{subfigure}
\hfill
\begin{subfigure}[t]{0.49\linewidth}
\includegraphics[width=\linewidth]{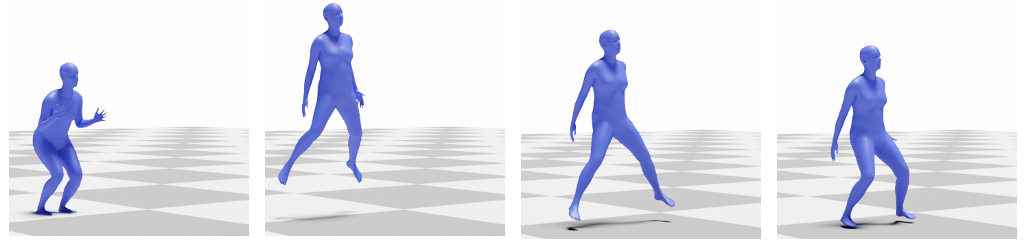}
\caption{Edit: \textit{As you jump, kick both legs out to the side.}}

\end{subfigure}

% Second Row
\begin{subfigure}[t]{0.49\linewidth}
\includegraphics[width=\linewidth]{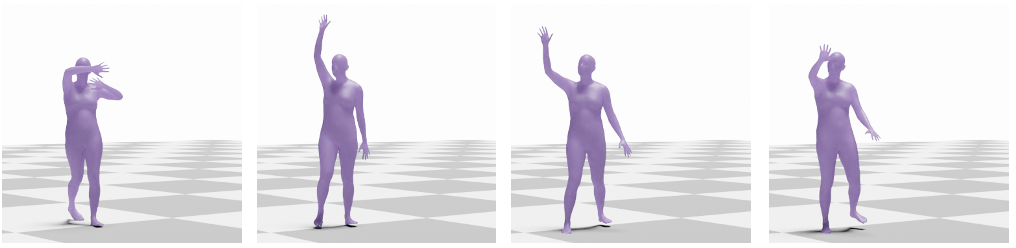}
\caption{Original Motion: \textit{A person is swinging their arms.}}

\end{subfigure}
\hfill
\begin{subfigure}[t]{0.49\linewidth}
\includegraphics[width=\linewidth]{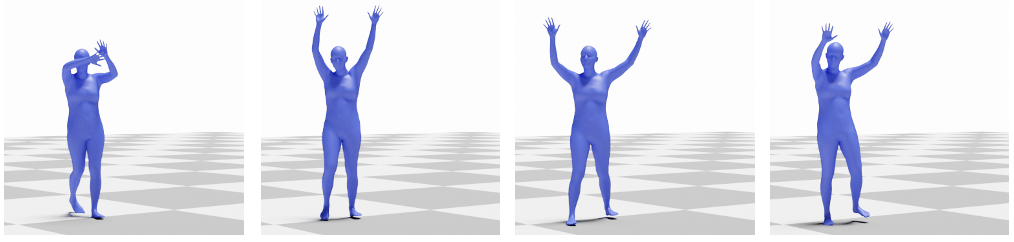}
\caption{Edit: \textit{Synchronize your arms.}}
\end{subfigure}
\caption{\textbf{Handling natural-language instructions.} Starting from a source motion (left column, in purple) and editing instruction (italicized), our system produces plausible motions (right column, blue) that preserve the structure of the original motion and abide by the editing instruction.}
\label{fig:edits_qual}
\end{figure*}

\subsection{Qualitative Evaluation}

We used our system to edit a variety of motions using natural language. In Figures~\ref{system} and~\ref{fig:edits_qual}, and our Supplemental Materials, we demonstrate how on a per-edit basis, the system can handle a range of editing intents, and produce a variety of motions that are faithful to the edit, preserve qualities of the original motion, and are visually plausible. Our primary goal, though, is to provide a conversational interface supporting iterative editing; we show results of iterative editing sessions in Figure~\ref{fig:banner} and in the accompanying video. In these examples, instructions are used to progressively refine the character's motion, break complex goals into step-by-step instructions, and also clarify or adjust editing intent during the refinement process. 

\subsection{Quantitative Evaluation}
\subsubsection{Experimental Setup}
To quantitatively evaluate our system, we compare its performance against two SOTA text-to-motion models: MDM~\cite{tevet2023human} and MoMask~\cite{guo2023momask}, using automated metrics and a user study. 

MDM cannot take source motions as input; therefore, to generate edited motions, we first write ten captions that are plentiful in MDM's training data (HumanML3D~\cite{Guo_2022_CVPR}) like kicking and throwing, e.g., ``A person is kicking with the right foot.'' Captions are fed into MDM to generate several motions to be \Xsrc. Next, we concatenate each source caption with different editing instructions, e.g., ``A person is kicking with the right foot. As you kick, raise your arms out to the side.'' Editing instructions were inspired by kinematic motion descriptions that appeared often in HumanML3D. For the baseline, MDM-Edit, we fix MDM's generation seed and rerun it on the concatenated caption. We compare MDM-Edit with our system's editing of \Xsrc using the same caption.

MoMask can perform mask-based editing, e.g., inpainting source motions within a specified mask given a new prompt, but cannot deduce mask frame indices. So, we generate a separate set of \Xsrc with MoMask, then employ our LLM-based parser to determine the frame(s) associated with different editing prompts. We rerun MoMask with the editing prompts using masks centered around these frames, producing baseline MoMask-Edit.

\subsubsection{User Study}
\label{subsec:humaneval}

\begin{figure}
\centering
\includegraphics[page=5,trim={0 925 0 450},clip,width=\linewidth]{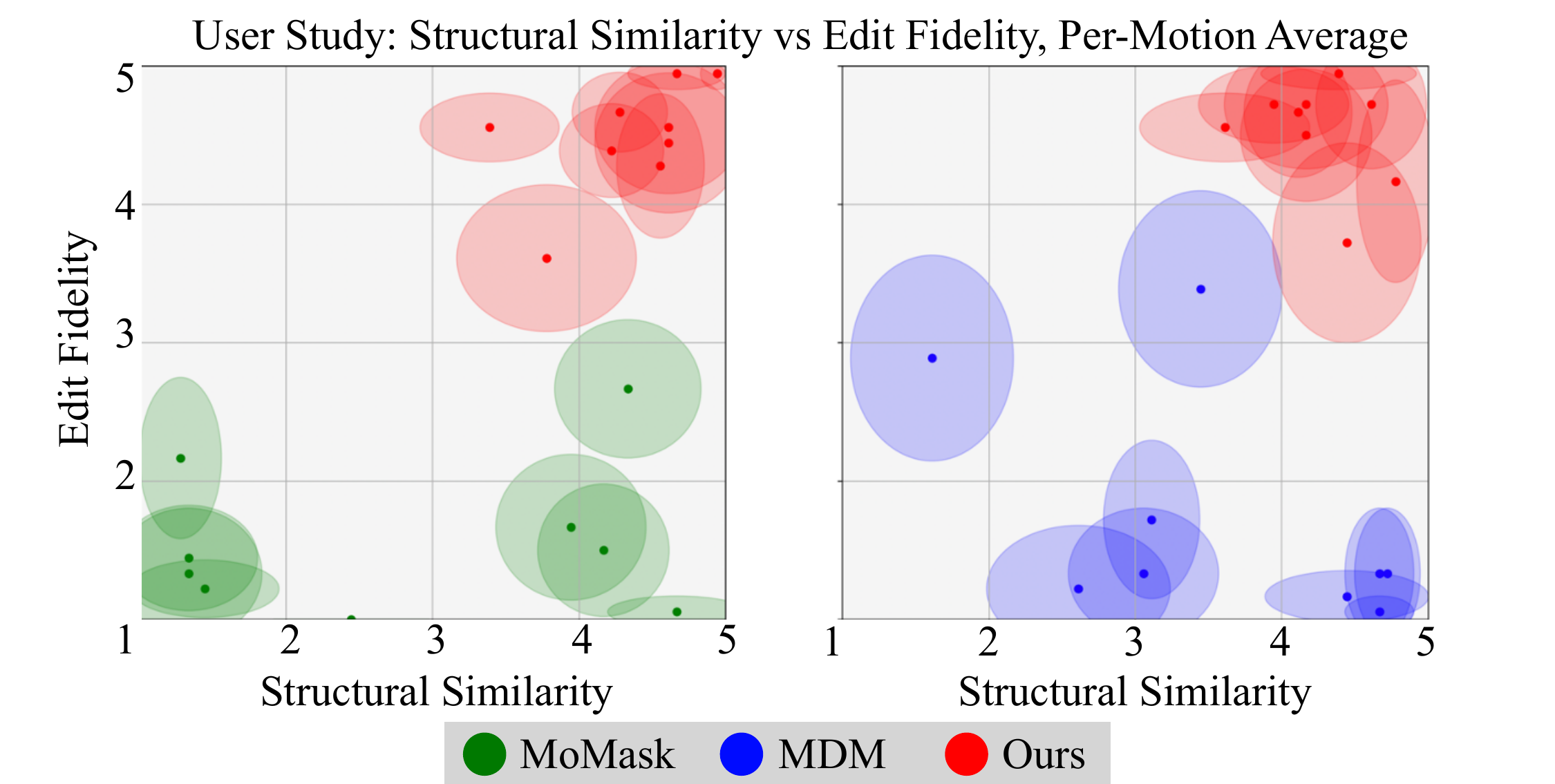}
\caption{\textbf{Per-motion average score for edit fidelity vs structural similarity} in our user study. MDM-Edit (blue) and MoMask-Edit (green) struggle to achieve a high score on both axes at the same time; high scores in structural similarity are often at the cost of edit fidelity. In contrast, our system (red) simultaneously scores high on both.}
\label{userstudygraph}
\end{figure}

% preserve the overall structure of . Source motions were retrieved from real mocap data, i.e., from AMASS, and from text-to-motion models, i.e., MDM. We also show examples of iterative editing.
We conduct a user study to compare a sample of edited motions. 19 users rated nine MDM-Edit motions versus our edited versions, and nine MoMask-Edit motions versus our edited versions. Users were asked to rate each result based on the overall quality of each motion, fidelity to the editing instruction, and structural similarity to the original motion from 1-5 (higher is better); details are in the Supplemental. 
We show average scores in Table~\ref{fig:user_study}. Our system's Fidelity
scores far exceed both baselines, and were judged better on structural similarity. 
On a per-motion level, we observe that though baselines can sometimes maintain structural similarity, they often struggle to simultaneously maintain Fidelity. In contrast, our motions score high on both axes; see Fig.~\ref{userstudygraph}.
% 2. We observe MDM can be good on either but not both. Sometimes it ignores the edit, and produces great new motion. (see graph)

%users were not professional animators but were familiar with computer graphics

\begin{table}[t!]
\centering
    \caption{\textbf{User study results.} 19 participants rated faithfulness of the edited motion to the instruction (Fidelity), preservation of the source motion's structure (StrucSim), and motion quality (Qual). We report average scores over all users and motions; we score much higher on StrucSim and Fidelity, and similarly on Qual. }
 \label{fig:user_study}
 \begin{tabular}{c c c c} 
 \toprule
  & StrucSim ($\uparrow$) & Fidelity ($\uparrow$) & Qual ($\uparrow$)  \\ 
 \midrule
 MDM-Edit & 2.77 & 1.56 & 3.79 \\
 Ours & 4.33 & 4.48 & 4.07 \\
 \midrule
 MoMask-Edit & 3.59 & 1.72 & 3.88 \\
 Ours & 4.25 & 4.52 & 3.77 \\
 \bottomrule
 \end{tabular}
\end{table}

\subsubsection{Metrics}

% To compare CMC and MDM-PE on a larger range of motion sequences, we scale our evaluation using automatic measures G-MPJPE and Fidelity-Auto on 50 edited motions. Further details on both measures are included in 5.2.1 and Appendix D.

We also evaluate our system against baselines using automated metrics for an additional 17 edited motion pairs for each baseline. We measure structural similarity using G-MPJPE, a common geometric distance metric in motion reconstruction. To measure edit fidelity, we author binary edit fidelity tests that use joint positions to assess whether changes fulfill the desired intent of a given MEO. We rate edit fidelity by the average number of tests passed (Fidelity-Auto). We measure quality using Frechet Distance to compare an empirical distribution against 1000 ground truth motions in the \textit{fairmotion}~\cite{gopinath2020fairmotion} geometric feature space ($FID_g$)~\cite{li2021ai}. See the Supplemental for more details.

Quantitative metrics reveal similar trends to our user study. Against MoMask-Edit, our edited motions score 140\% higher on Fidelity-Auto (0.88 versus 0.6), and are structurally more similar to MoMask-Source. We show similar improvement over MDM-Edit--see Table~\ref{fig:more_quant}. 

We do not compare the motion quality of our system vs MDM-Edit or MoMask-Edit quantitatively here; all are editing motions that have been generated by MDM/MoMask, which already have some deviation from ground-truth human motions that would affect overall quality scores of their edited versions.

\begin{table}[t!]
\caption{\textbf{Quantitative evaluation with automated metrics.} Both versus MoMask-Edit and MDM-Edit, our system scores more favorably on edit fidelity and G-MPJPE. We find our evaluation to be statistically significant with pairwise comparisons using Wilcoxon’s signed rank test. Ours vs MDM: p<0.03,Z=24 for fidelity and p<0.0003,Z=10 for GMPJPE. Ours vs MoMask: p<0.02,Z=10 for fidelity and p<0.0002,Z=8 for GMPJPE.
% Additional automatic quantitative scoring of larger datasets.}
}
 \label{fig:more_quant}
\begin{tabular}{l|cc|cc}
\toprule
                     & \multicolumn{1}{c}{MDM-Edit} & \multicolumn{1}{c}{Ours} & \multicolumn{1}{|c}{MoMask-Edit} & \multicolumn{1}{c}{Ours} \\
\midrule

Fidelity ($\uparrow$)  & 0.588                       & 0.82                    & 0.6                       & 0.882                    \\
G-MPJPE ($\downarrow$) & 0.247                       & 0.08                    & 0.181                       & 0.063                   \\
\bottomrule
\end{tabular}
\end{table}

% \begin{table}[t!]
% \caption{\textbf{Quantitative evaluation with automated metrics.} Our system produces motions that score higher on edit fidelity, and more favorably on G-MPJPE, our measure of structural similarity.
% % Additional automatic quantitative scoring of larger datasets.}
% }
%  \label{fig:more_quant}
% \begin{tabular}{l|cc|cc}
% \toprule
%                      & \multicolumn{1}{c}{MDM-PE} & \multicolumn{1}{c}{Ours} & \multicolumn{1}{|c}{MDM-PE} & \multicolumn{1}{c}{Ours} \\
% \midrule
%                      & \multicolumn{2}{c}{\textit{\small User Study}}                       & \multicolumn{2}{|c}{\textit{\small 
%  Extra}}                            \\

% Fidelity ($\uparrow$)  & 0.29                       & 0.93                    & 0.64                       & 0.91                    \\
% G-MPJPE ($\downarrow$) & 0.25                       & 0.10                    & 0.17                       & 0.06                   \\
% \bottomrule
% \end{tabular}
% \end{table}

\subsubsection{Execution Engine Ablation Study}
\label{subsec:ablation}

We measure motion quality over an ablation of our execution engine. We start with 100 real mocap sequences in AMASS (AMASS-Source). We pair each with 1-3 MEOs and edit AMASS-Source using ablated versions of the engine: 1) ENG, our proposed engine, 2) ENG-SS, where diffusion model \textit{G} is trained to infill the entire body instead of our two-stage process, and 3) ENG-Interp, where we use spline interpolation instead of \textit{G}. We compare $FID_g$ for each engine's generations.

ENG produces motion distributions that more closely match those of source motions. Edited motions should preserve the overall structure of the source, so we expect $FID_{g}$ of edited motions to \textit{match} $FID_{g}$ of AMASS-Source, rather than improve upon it. Indeed, AMASS-Source scores 4.33 and ENG only observes marginal increase to 4.95. Ablations degrade the $FID_g$ score: ENG-SS drops the $FID_g$ to 5.25 and as we expect,  ENG-Interp's spline interpolation produces the least ``human-like" results with $FID_g$ of 8.05.

\section{Discussion and Limitations}

\paragraph{\textbf{Limitations}} MEOs are limited to kinematic constraints; physics-informed edits like, ``jump more forcefully'' are not handled by our system. Extending the execution engine to support these edits are exciting future directions. In our system, source motion context and keyframes act as a condition for the diffusion model, but should not necessarily be considered as ``hard" spatiotemporal constraints, e.g., editing of joint positions can result in extra displacement of a joint, which in turn should require more transition time to avoid velocity inconsistencies. We are eager to explore methods to improve motion quality and make the execution engine more robust to such input. Currently, our system's frame-picking is largely based on joint extrema; more sophisticated methods for motion understanding~\cite{endo2023humanmotionqa} could make this more flexible. 

In conclusion, we have demonstrated a system for editing motions with text, by first translating text instructions into keyframe-like ``constraints". Our system can iteratively edit motions from a variety of sources: mocap datasets~\cite{AMASS:ICCV:2019}, modern generative models~\cite{tevet2023human, guo2023momask}, extracted from video~\cite{Wang_2023_CVPR}, etc.
We are excited about expanding the scope user inputs to the system to more than just text-based instruction and adding new MEO operators, which we believe can specify many edits, e.g., stylistic changes and physically-informed objectives. 
Extending the system in this manner would provide new ways for users to direct characters.

\begin{acks}
Purvi Goel is supported by a Stanford Interdisciplinary Graduate Fellowship. Kuan-Chieh Wang was supported by Stanford Wu-Tsai Human Performance Alliances while at Stanford University. We thank the anonymous reviewers for constructive feedback; Vishnu Sarukkai, Sarah Jobalia, Sofia Di Toro Wyetzner for proofreading; Haotian Zhang, David Durst, and James Hong for helpful discussions. Our codebase was built with invaluable help from James Burgess. 
\end{acks}

%%%%%%%%% REFERENCES
{
    \small
    \bibliographystyle{ACM-Reference-Format}
    \bibliography{main}
}

\end{document}